\documentclass{article}

\usepackage{microtype}
\usepackage{graphicx}
\usepackage{subfigure}
\usepackage{booktabs}
\usepackage{pdfpages}
\usepackage{multirow}
\usepackage{mathtools}
\DeclarePairedDelimiter\ceil{\lceil}{\rceil}

\usepackage{hyperref}
\usepackage{bm}
\usepackage{amssymb,amsmath,graphicx,bbm,color,booktabs}
\usepackage{amsopn}

\newcommand{\ve}{\bm{e}}

\newcommand{\vs}{\bm{s}}       \newcommand{\vsh}{\hat{\bm{s}}}        \newcommand{\sh}{\hat{s}}
               
\newcommand{\vu}{\bm{u}}               
\newcommand{\vv}{\bm{v}}               
               
\newcommand{\vx}{\bm{x}}               
               
\newcommand{\vz}{\bm{z}}               

\newcommand{\R}{\mathbb{R}}

\renewcommand{\S}{\mathcal{S}}

\renewcommand{\eqref}[1]{Eq.~(\ref{#1})}

\usepackage[accepted]{icml2020}

\icmltitlerunning{Voice Separation with an Unknown Number of Multiple Speakers}

\newcommand{\ours}{\textsc{MulCat}~}

\begin{document}

\twocolumn[
\icmltitle{Voice Separation with an Unknown Number of Multiple Speakers}

\begin{icmlauthorlist}
\icmlauthor{Eliya Nachmani}{fair,tau}
\icmlauthor{Yossi Adi}{fair}
\icmlauthor{Lior Wolf}{fair,tau}
\end{icmlauthorlist}

\icmlaffiliation{fair}{Facebook AI Research}
\icmlaffiliation{tau}{Tel-Aviv University}

\icmlcorrespondingauthor{Eliya Nachmani}{enk100@gmail.com}
\icmlcorrespondingauthor{Yossi Adi}{yossiadidrum@gmail.com}
\icmlcorrespondingauthor{Lior Wolf }{liorwolf@gmail.com}

\icmlkeywords{Machine Learning, source separation, speaker separation}
\vskip 0.3in 
]
\printAffiliationsAndNotice{}

\begin{abstract}
We present a new method for separating a mixed audio sequence, in which multiple voices speak simultaneously. The new method employs gated neural networks that are trained to separate the voices at multiple processing steps, while maintaining the speaker in each output channel fixed. A different model is trained for every number of possible speakers, and the model with the largest number of speakers is employed to select the actual number of speakers in a given sample. Our method greatly outperforms the current state of the art, which, as we show, is not competitive for more than two speakers. 
\end{abstract}

\section{Introduction}

The ability to separate a single voice from the multiple conversations occurring concurrently forms a challenging perceptual task~\cite{capon1969high,frost1972algorithm}. The ability of humans to do so has inspired many computational attempts, with much of the earlier work focusing on multiple microphones and unsupervised learning, e.g., the independent component analysis approach~\cite{hyvarinen2000independent}.

In this work, we focus on the problem of supervised voice separation from a single microphone, which has seen a great leap in performance following the advent of deep neural networks~\cite{hershey2016deep,luo2018tasnet}. In this ``single-channel source separation'' problem, given a dataset containing both the mixed audio and the individual voices, one trains to separate a novel mixed audio that contains multiple unseen speakers. 

The current leading methodology is based on an overcomplete set of linear filters, and on separating the filter outputs at every time step using a mask for two speakers, or a multiplexer for more speakers~\cite{luo2018tasnet,luo2019conv,zhang2020furcanext}. The audio is then reconstructed from this partial representations. Since the order of the speakers is considered arbitrary (it is hard to sort voices), one uses a permutation invariant loss during training, such that the permutation that minimizes the loss is considered. 

The need to work the aforementioned partial representations, which becomes more severe as the number of voices to be separated increases, is a limitation of this masking-based method, since the mask needs to extract and suppress more from the representation as the number speakers increases. In this work, we, therefore, set out to build a mask-free method. The method employs a sequence of RNNs that are applied to the audio. As we show, it is beneficial to evaluate the error after each RNN, obtaining a compound loss that reflects the reconstruction quality after each layer. 

The RNNs are bi-directional. Each RNN block is built with a specific type of residual connection, where two RNNs run in parallel. The output of each layer is the concatenation of the element-wise multiplication of the two RNNs together with the layer input that undergoes a bypass (skip) connection. 

Unlike separating known sources~\cite{defossez2019music} in this case, the outputs are given in a permutation invariant fashion, hence, voices can switch between output-channels, especially during transient silence episodes. In order to tackle this, we propose a new loss that is based on a voice representation network that is trained on the same training set. The embedding obtained by this network is then used to compare the output voice to the voice of the output channel. We demonstrate that the loss is effective, even when adding it to the baseline method. An additional improvement, that is effective also for the baseline methods, is obtained by starting the separation from multiple locations along the audio file and averaging the results. 

Similar to the state of the art methods, we train a single model for each number of speakers. The gap in performance of the obtained model in comparison to published methods increases as the number of speaker increases, and one can notice that the performance of our method degrades gradually, while the baseline methods show a sharp degradation as the number of speakers increases.

To support the possibility of working with an unknown number of speakers, we opt for a learning-free solution and select the number of speakers by running an activity detector on its output. This simple method is able to select the correct number of speakers in the vast majority of the cases and leads to our method being able to handle an unknown number of speakers.

Our contributions are: (i) a novel audio separation model that employs a specific RNN architecture, (ii) a set of losses for effective training of voice separation networks, (iii) performing effective model selection in the context of voice separation with an unknown number of speakers, and (iv) state of the art results that show a sizable improvement over the current state of the art in an active and competitive domain.

\section{Model}
In the problem of single-channel source separation, the goal is to estimate $C$ different input sources $\vs_j \in \R^T$, where $j \in [1, \cdots, C]$, given a mixture $\vx = \sum_{i=1}^C c_i \cdot s_i$, where $c_i$ is a scaling factor. The input length, T, is not a fixed value, since the input utterances can have different durations. In this work, we focus on the supervised setting, in which we are provided with a training set $\S = \{\vx_i, (\vs_{i, 1}, \cdots, \vs_{i, C})\}_{i=1}^n$, and our goal is learn the model that given an unseen mixture $\vx$, outputs $C$ estimated channels $\sh = (\vsh_{1},  \cdots, \vsh_{C})$ that maximize the scale-invariant source-to-noise ratio (SI-SNR) (also known as the scale-invariant signal-to-distortion ratio, SI-SDR for short), between the predicted and the target utterances. More precisely, since the order of the input sources is arbitrary and since the summation of the sources is order invariant, the goal is to find $C$ separate channels $s$ that maximize the SI-SNR to the ground truth signals, when considering the reorder channels $(\vsh_{\pi(1)}, \cdots, \vsh_{\pi(C)})$ for the optimal permutation $\pi$.

\subsection{Model Description}

\begin{figure*}[t]
   \centering
   \includegraphics[page=1,width=1.\textwidth]{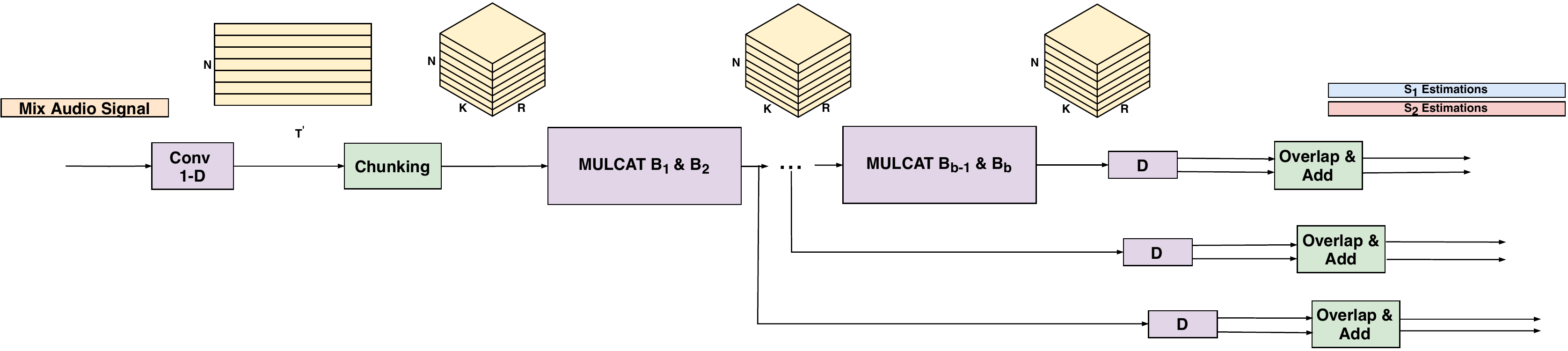}
    \caption{The architecture of our network. The audio is being convolved with a stack of 1D convolutions and reordered by cutting overlapping segments of length $K$ in time, to obtain a 3D tensor. 
In our method, the RNN blocks are of the type of multiply and add. After each pair of blocks, we apply a convolution $D$ to the copy of the activations, and obtain output channels by reordering the chunks and then using the overlap and add operator.}
 \label{fig:arch}
\end{figure*}

\begin{figure*}[t]
   \centering
    \includegraphics[page=1,width=1\textwidth]{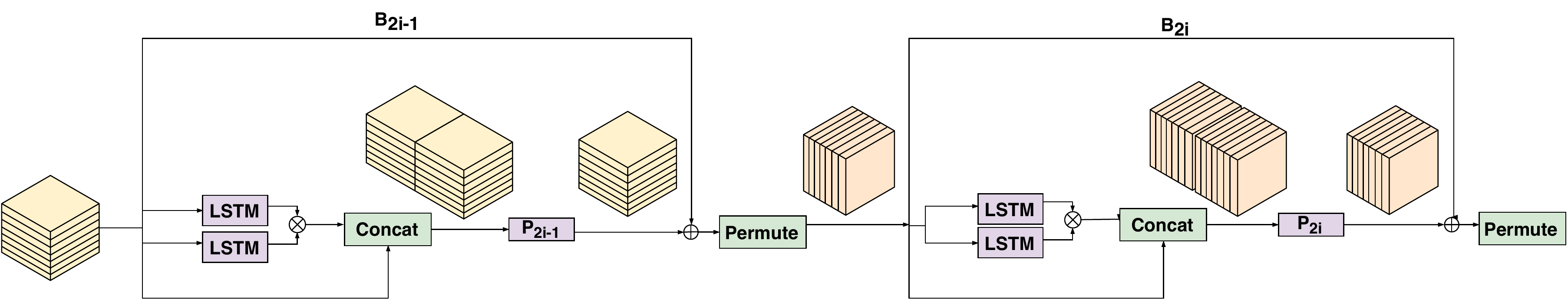}
    \caption{The multiply and concat (\ours) block. In the odd blocks, the 3D tensor obtained from chunking is fed into two different bi-directional LSTMs that operate along the second dimension. The results are multiplied element-wise, followed by a concatenation of the original signal along the third dimension. A learned linear projection along this dimension is then applied to obtain a tensor of the same size of the input. In the even blocks, the same set of operations occur along the chunking axis.}
 \label{fig:multcat}
\end{figure*}

The proposed model, depicted in Figure~\ref{fig:arch}, is inspired by the recent advances in speaker separation models~\cite{luo2018tasnet, luo2019conv}. The first steps of processing, including the encoding, the chunking, and the two bi-directional RNNs on the tensor that is obtained from chunking are similar. However, our RNNs contain dual heads, we do not use masking, and our losses are different.

First, an encoder network, $E$, gets as input the mixture waveform $\vx\in \R^T$ and outputs a $N$-dimensional latent representation $\vz$ {of size $T^{`} = (2T/L)-1$, where $L$ is the encoding compression factor. This results in $\vz \in \R^{N \times T^{`}}$},
\begin{equation}
\label{eq:enc_net}
\vz = E(\vx)
\end{equation}
Specifically, $E$ is a 1-D convolutional layer with a kernel size $L$ and a stride of $L/2$, followed by a ReLU non-linear activation function. 

The latent representation $\vz$ is then divided into {$R=\ceil{2T^{`}/K}+1$} overlapping chunks of length $K$ and hop size $P$, denoted as $\vu_r \in \R^{N\times K}$, where $r \in [1, \cdots , R ]$. All chunks are then being concatenated along the singleton dimensions and we obtain a 3-D tensor $\vv = [\vu_1, \dots ,\vu_R] \in \R^{N\times K\times R}$. 

Next, similar to~\cite{luo2019dual}, $\vv$ is fed into the separation network $Q$, which consists of $b$ RNN blocks. The odd blocks $B_{2i-1}$ for $i=1, \dots,b/2$ apply the RNN along the time-dependent dimension of size $R$. The even $B_{2i}$ blocks are applied along the chunking dimension of size $K$. Intuitively, processing the second dimension yields a short-term representation, while processing the third dimension produce long-term representation.

At this point, our method diverges from~\cite{luo2019dual}, since our RNN blocks contain the \ours block with two sub-networks and a skip connection. Consider, for example, the odd blocks $B_i$, $i=1,3,\dots,b-1$. We employ two separate bidirectional LSTM, denoted as $M^1_i$ and $M^2_i$, element wise multiply their outputs, and finally concatenate the input to produce the module output. 

\begin{equation}
    B_i(\vv) = P_i([M^1_i(\vv)\odot M^2_i(\vv),\vv])
\end{equation}
where $\odot$ is the element wise product operation, and $P_i$ is a learned linear project that brings the dimension of the result of concatenating the product of the two LSTMs with the input $\vv$ back to the dimension of $\vv$. A visual description of a pair of blocks is given in Figure~\ref{fig:multcat}.

In our method, we employ a multi-scale loss, which requires us to reconstruct the original audio after each pair of blocks. The 3D tensor undergoes the PReLU non-linearity~\cite{he2015delving} with parameters initialized at 0.25. Then, a $1 \times 1$ convolution with $CR$ output channels, denoted as $D$. The resulting tensor of size $N\times K\times CR$ is divided into $C$ tensors of of size $N\times K\times R$ that would lead to the $C$ output channels. Note that the same PReLU parameters and the same decoder $D$ are used to decode the output of every pair of \ours blocks.

In order to transform the 3D tensor back to audio, we employ the overlap-and-add operator to the $R$ chunks~\cite{rabiner1975theory}. The operator, which inverts the chunking process, adds overlapping frames of the signal after offsetting them appropriately by a step size of $L/2$ frames.

\subsection{Training Objective}
Recall that since the identity of the speakers is unknown, our goal is to find $C$ separate channels $\sh$ that maximize the SI-SNR between the predicted and target signals. Formally, the SI-SNR is defined as
\begin{equation}
\label{eq:sisnr}
\text{SI-SNR}(\vs, \vsh) = 10\log_{10} \frac{\|\tilde{\vs_i}\|^2}{\|\tilde{\ve_i}\|^2}
\end{equation}
where, $\tilde{\vs_i} = \frac{\langle\vs_i, \vsh_i\rangle\vs_i}{\|\vs_i\|^2}$, and $\tilde{\ve_i} = \vsh_i - \tilde{\vs_i}$. 

Since the channels are unordered, the loss is computed for the optimal permutation $\pi$ of the $C$ different output channels and is given as:
\begin{equation}
\label{eq:optimal}
\ell(s, \sh) = -\max_{\pi \in \Pi_C}~\frac{1}{C}\sum_{i=1}^C \text{SI-SNR}(\vs_i \vsh_{\pi(i)})
\end{equation}
where $\Pi_C$ is the set of all possible permutations of $1\dots C$. 
The loss $\ell(s,\sh)$ is often denoted as the utterance level permutation invariant training (uPIT)~\cite{kolbaek2017multitalker}. 

As stated above, the convolution $D$ is used to decode after every even \ours block, allowing us to apply the uPIT loss multiple times along the decomposition process. Formally, our model outputs $b/2$ groups of output channels $\{\sh_j\}_{j=1}^{b/2}$ and we consider the loss
\begin{equation}
\begin{split}
\label{eq:obj}
\ell(s, \{\sh_j\}_{j=1}^{b/2}) = \frac{1}{b}\sum_{j=1}^{b/2} \ell(\vs, \vsh_{j}) \\
\end{split}
\end{equation}
Notice that the permutation of $\pi$ the output channels may be different between the components of this loss.

\paragraph{Speaker Classification Loss.} 
A common problem in source separation is forcing the separated signal frames belonging to the same speaker to be aligned with the same output stream. Unlike the Permutation Invariant Loss (PIT)~\cite{yu2017permutation} which is applied to each input frame independently, the uPIT is applied to the whole sequence at once. This modification greatly improves the amount of occurrences in which the output is flipped between the different sources. However, according to our experiments (See Section~\ref{sec:exp}) this is still a far from being optimal. 

To mitigate that, we propose to add an additional loss function which imposes a long term dependency on the output streams. For this purpose, we use a speaker recognition model that we train to identify the persons in the training set. Once this neural network is trained, we minimize the L2 distance between the network embeddings of the predicted audio channel and the corresponding source. 

\begin{figure}[t]
   \centering
       \includegraphics[page=1,width=.5\textwidth]{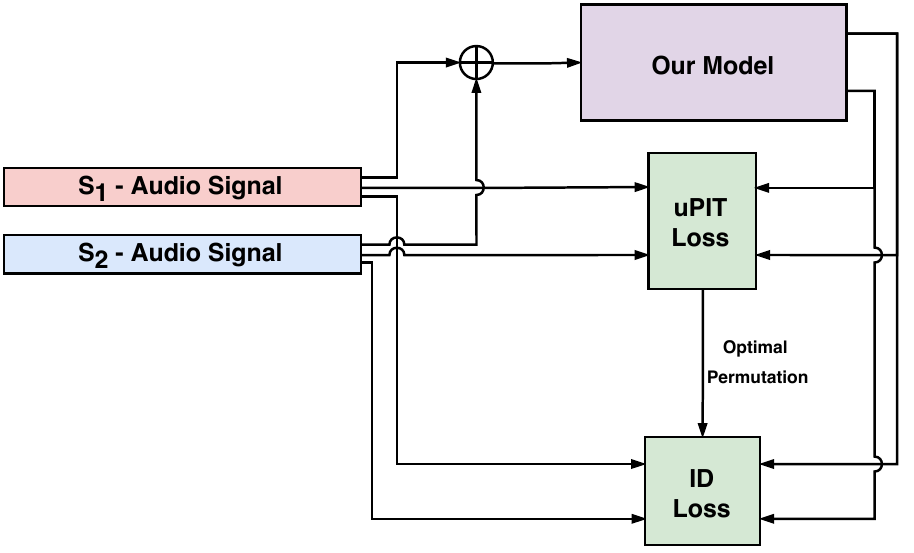}
    \caption{The training losses used in our method, shown for the case of $C=2$ speakers. The mixed signal $x$ combines the two input voices $s_1$ and $s_2$. Our model then separates to create two output channels $\hat s_1$ and $\hat s_2$. The permutation invariant SI-SNR loss computes the SI-SNR between the ground truth channels and the output channels, obtained at the channel permutation $\pi$ that minimizes the loss. The identity loss is then applied to the matching channels, after they have been ordered by $\pi$.}
 \label{fig:vgg_loss}
\end{figure}

As the speaker recognition model, we use the VGG11 network~\cite{simonyan2014very} trained on the power spectrograms (STFT) obtained from 500 ms of audio. Denote the embedding obtained from the penultimate layer of the trained VGG network by $G$. We used it in order to compare segments of length 500 ms of the ground truth audio $\vs_i$ with the output audio $\hat \vs_{\pi(i)}$, where $\pi$ is the optimal permutation obtained from the uPIT loss, see Fig.~\ref{fig:vgg_loss}.

Let $\vs_i^j$ be the $j$-th segments of length 500 ms obtained by cropping audio sequence $\vs_i$, and similarly $\hat \vs_i^j$ for $\vs_i$. The identity loss is give by
\begin{equation}
\label{eq:vgg}
\ell_{ID}(\vs,\hat\vs) = \frac{1}{C|J(s)|}\sum_{i=1}^C\sum_{j=1}^{J(s)} MSE(G(F(\vs_i^j)), G(F(\hat \vs_i^j)))
\end{equation}
where $J(\vs)$ is the number of segments extracted from $\vs$ and $F$ is a differential STFT implementation, i.e., a network implementation of STFT that allows us to back-propagate the gradient though it.

\subsection{Model Selection}
\label{sec:select}

We train a different model for each number $C$ of audio components in the mix. This allows us to directly compare with the baseline methods. However, in order to apply the method in practice, it is important to be able to select the number of speakers. While it is possible to train a classifier to determine $C$ given a mixed audio, we opt for a non-learned solution in order to avoid biases that arise from the distribution of data and to promote solutions in which the separation models are not detached from the selection process. 

The procedure we employ is based on activity detection algorithm, in which we compute the average power of each output channel and verify that it is above a predefined threshold~\footnote{we calibrated the threshold on the validation set.}. Starting from the model that was trained on the dataset with the largest number of speakers $C$, we apply the speech detector to each output channel. If we detect silence (no-activity) in one of the channels, we move to the model with $C-1$ output channels and repeat the process until all output channels contain speech.

As can be seen in our experiments, this selection procedure is relatively accurate and leads to results with an unknown number of speakers that are only moderately worse than the results when this parameter is known.

\section{Experiments}
\label{sec:exp}

For research purposes only, we use the WSJ0-2mix and WSJ0-3mix datasets \cite{hershey2016deep} and we further expand the WSJ-mix dataset to four and five speakers and introduce WSJ0-4mix and WSJ0-5mix datasets. All of the aforementioned datasets are based on the WSJ0 corpus~\cite{garofolo1993csr}. We use the same procedure as in \cite{hershey2016deep}, i.e. we use 30 hours of speech from the training set si$\_$tr$\_$s to create the training and validation sets. The four and five speakers were randomly chosen and combined with random SNR values between $0-5$ dB. The test set is created from  si$\_$et$\_$s and si$\_$dt$\_$s with 16 speakers, that differ from the speakers of the training set. For research purposes only, WSJ0-4mix and WSJ0-5mix datasets creation scripts are available as supplementary. A separate model is trained for each dataset, with the corresponding number of output channels. Sample results and db creation scripts can be found under the following link: \href{https://enk100.github.io/speaker_separation}{https://enk100.github.io/speaker$\_$separation}

\noindent{\bf Implementation details} We choose hyper parameters based on the validation set.  The input kernel size $L$ was $8$ (except for the experiment where we vary it) and the number of the filter in the preliminary convolutional layer was $128$. We use an audio segment of four seconds long sampled at 8kHz. The architecture uses $b=6$ blocks of \ours, where each LSTM layer contains $128$ neurons. We multiply the IDloss with $0.001$ when combined the uPIT loss. The learning rate was set to $5e-4$, which was multiplied by $0.98$ every two epoches. The ADAM optimizer \cite{kingma2014adam} was used with batch size of $2$. For the speaker model, we extract the STFT using a window size of 20ms with stride of 10ms and Hamming window.

In order to evaluate the proposed model, we report the scale-invariant signal-to-noise ratio improvement (SI-SNRi) score on the test set, computed as follows,
\begin{equation}
    \text{SI-SNRi}(\vs, \vsh, \vx) = \frac{1}{C} \sum_{i=1}^{C} \text{SI-SNR}(\vs_i, \vsh_i) - \text{SI-SNR}(\vs_i, \vx)
\end{equation}

We compare with the following baseline methods: ADANet~\cite{luo2018speaker}, DPCL++~\cite{isik2016single}, CBLDNN-GAT~\cite{li2018cbldnn}, TasNet~\cite{luo2018tasnet}, the Ideal Ratio Mask (IRM), the Ideal Binary Mask (IBM), ConvTasNet~\cite{luo2019conv}, FurcaNeXt~\cite{zhang2020furcanext}, and DPRNN~\cite{luo2019dual}. Similarly to~\citet{luo2019conv}, for the IRM and IBM we use a window size of 32ms, hop length of 8ms, and 2048 FFT bins. Prior work often reports the signal-to-distortion ratio (SDR). However, recent studies have argued that the above mentioned metric has been improperly used due to its filter dependence and may result in misleading findings~\cite{le2019sdr}. 

The results are reported in Table~\ref{tab:main}. Each column depicts a different dataset, where the number of speakers $C$ in the mixed signal $x$ is different. The model used for evaluating each dataset is the model that was trained to separate the same number of speakers. As can be seen, the proposed model is superior to previous methods by a sizable margin, in all four datasets.

\begin{table}[t]
    \centering
    \caption{Performance of various models as a function of the number of speakers. Starred results (*) mark our training, using published code by the method's authors. The other baselines are obtained from the respective papers.}
    \label{tab:main}
	\resizebox{\columnwidth}{!}{
	\begin{tabular}{lccccc}
    \toprule
Model& $\#$params& 2spk&	3spk&	4spk&	5spk\\
\midrule
ADANet		& 9.1M  & 10.5	&	9.1	  &	-		&	-\\
DPCL++		& 13.6M & 10.8	&	7.1	  &	-		&	-\\
CBLDNN-GAT	& 39.5M	& 11		&	-	  &	-		&	-\\
TasNet		& 32.0M	& 11.2	&	-	  &	-		&	-\\
IBM			&	-	& 13.0	&	12.8  &	10.6		&	10.3\\
IRM			&	-	& 12.7	&	12.5  &	9.8 		&	9.6\\
ConvTasNet	& 5.1M	& 15.3	&	12.7  &	8.5*		&	6.8*\\
FurcaNeXt	& 51.4M	& 18.4	&	-	  &		- 	& 	-\\
DPRNN		& 3.6M	& 18.8	&	14.7* &	10.4*	&	8.7*\\
Ours		& 7.5M	&	{\bf 20.1}&	{\bf 16.9}&	{\bf 12.9}&	{\bf 10.6}\\
\bottomrule
\end{tabular}}
\end{table}

In order to understand the contribution of each of the various components in the proposed method, we conducted an ablation study. (i) We replace the \ours block with a conventional LSTM (``-gating''); (ii) we train with a permutation invariant loss that is applied only at the final output (``-multiloss'') of the model; and (iii) we train with and without the identity loss (``-IDloss''). 

First, we analyzed the importance of each loss term to the final model performance. Table~\ref{tab:ablation} summarizes the results. As can be seen, each of aforementioned components contributes to the performance gain of the proposed method, with the multi-layer loss being more dominant than the others. Adding the identity loss to the DPRNN model also yields a performance improvement. We would like to stress that not only being different in the multiply and concat block, the identity loss and the multiscale loss, our method does not employ a mask when performing separation and instead directly generates the separated signals.

\begin{table}[t]
    \centering
    \caption{Ablation analysis where we take out the two LSTM structures and replace them with a single one (-gating), remove the multiloss (-multiloss), or remove the speaker identification loss (-IDloss). We also present the results of adding the identification loss to the baseline DPRNN method. The DPRNN results are based on our training, using the authors' published code.}
    \label{tab:ablation}
    \begin{tabular}{@{}l@{~}c@{~~}c@{~~}c@{~~}c@{}}
    \toprule
Model&	2spk&	3spk&	4spk&	5spk\\
\midrule
DPRNN&	18.08 &	14.72&	10.37&	8.65\\
DPRNN + IDloss& 18.42 &	14.91&	11.29&	9.01\\
Ours-gating-multiloss-IDloss&	19.02&	14.88&	10.76&	8.42\\
Ours-gating-IDloss&	19.30&	15.60&	11.06&	8.84\\
Ours-gating		  &	19.42&	15.73&	11.22&	8.95\\
Ours-multiloss-IDloss&	18.84&	13.73&	10.40&	8.65\\
Ours-multiloss		 &	18.93&	13.86&	10.54&	8.75\\
Ours-IDloss&	19.76&	16.63&	12.60&	10.20\\
Ours&	\bf 20.12&	\bf 16.85&	\bf 12.88&	\bf 10.56\\
\bottomrule
\end{tabular}
\end{table}

\begin{figure}[t]
   \centering
       \includegraphics[page=1,width=.5\textwidth]{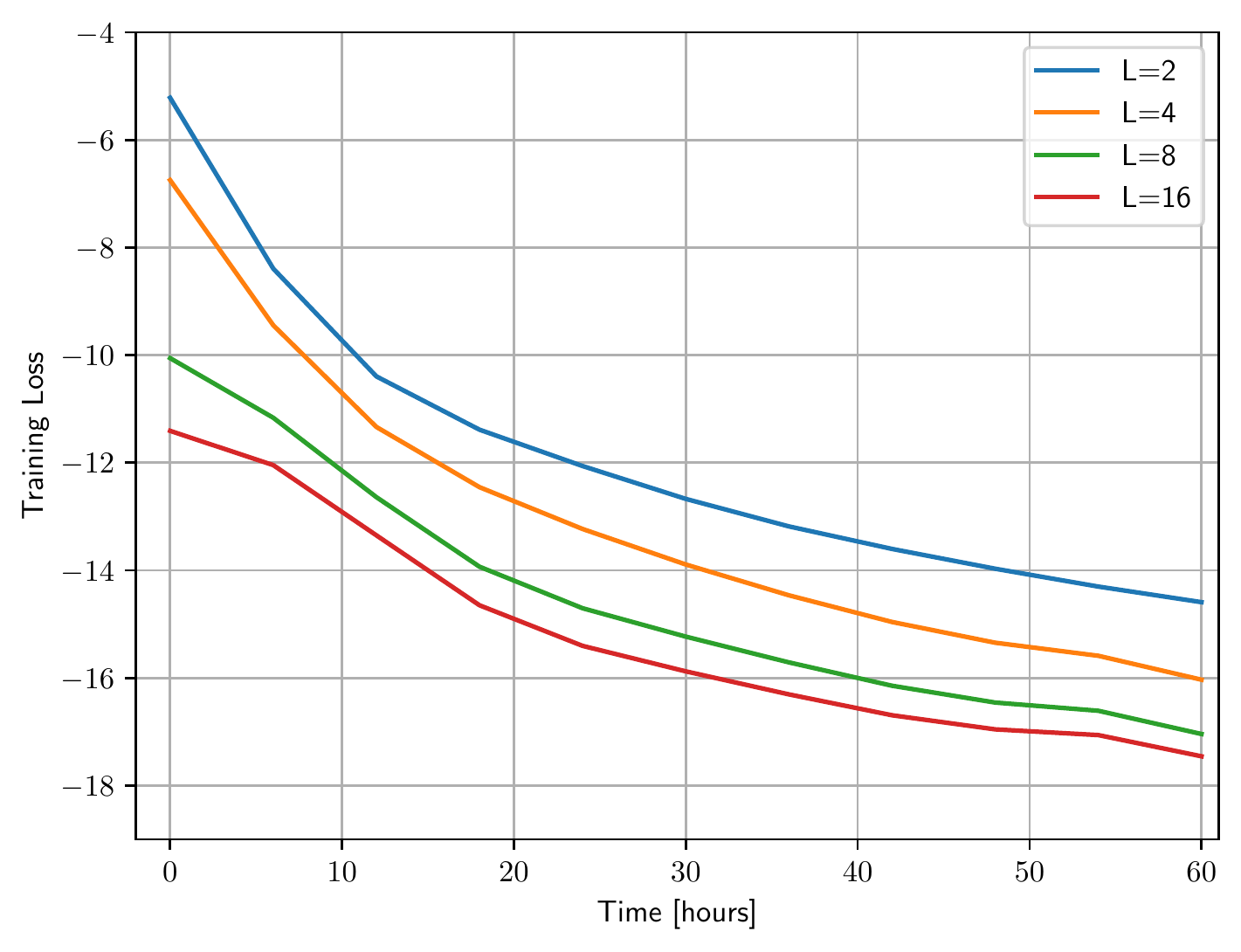}
    \caption{Training curves of our model for various kernel sizes $L=2,4,8,16$. Our model train faster with larger kernel size.}
 \label{fig:convergence}
\end{figure}

Recent studies pointed out the importance of choosing small kernel size for the encoder~\cite{luo2019dual}. In ConvTasNet the authors suggest that kernel size $L$ of 16~\cite{luo2019conv} performs better than larger ones, while the authors of DPRNN~\cite{luo2019dual} advocate for an even smaller size of $L=2$. Table~\ref{tab:funcofL} shows that unlike DPRNN, the performance of our model is not harmed by larger kernel sizes. Figure~\ref{fig:convergence} depicts the convergence rates of our model for various $L$ values for the first 60 hours of training. Being able to train with kernels with $L>2$ {\color{black}leads to faster convergence to results at the range of recently published methods}.

\begin{table}[t]
    \centering
    \caption{Performance of three types of models as a function of the kernel size. Our model does not suffer from changing the kernel size. (Only the last row is based on our runs).}
    \label{tab:funcofL}
    \begin{tabular}{lcccc}
    \toprule
Model&	L=2&	L=4&	L=8&	L=16\\
\midrule
ConvTasNet&	-&	-	&-	&15.3\\
DPRNN&	18.8&	17.9&	17.0&	15.9\\
Ours&	18.94&	19.91&	19.76&	18.16\\
\bottomrule
\end{tabular}
\end{table}

Lastly, we explored the effect of the identity loss. Recall that the identity loss is meant to reduce the frequency in which an output channel switches between the different speaker identities. In order to measure the frequency of this event, we have separated the audio into sub-clips of length 0.25sec and tested the best match, using SI-SNR, between each segment and the target speakers. If the matching switched from one voice to another, we marked the entire sample as a switching sample. 

\begin{figure}[t]
  \centering
  \includegraphics[page=1,width=.5\textwidth]{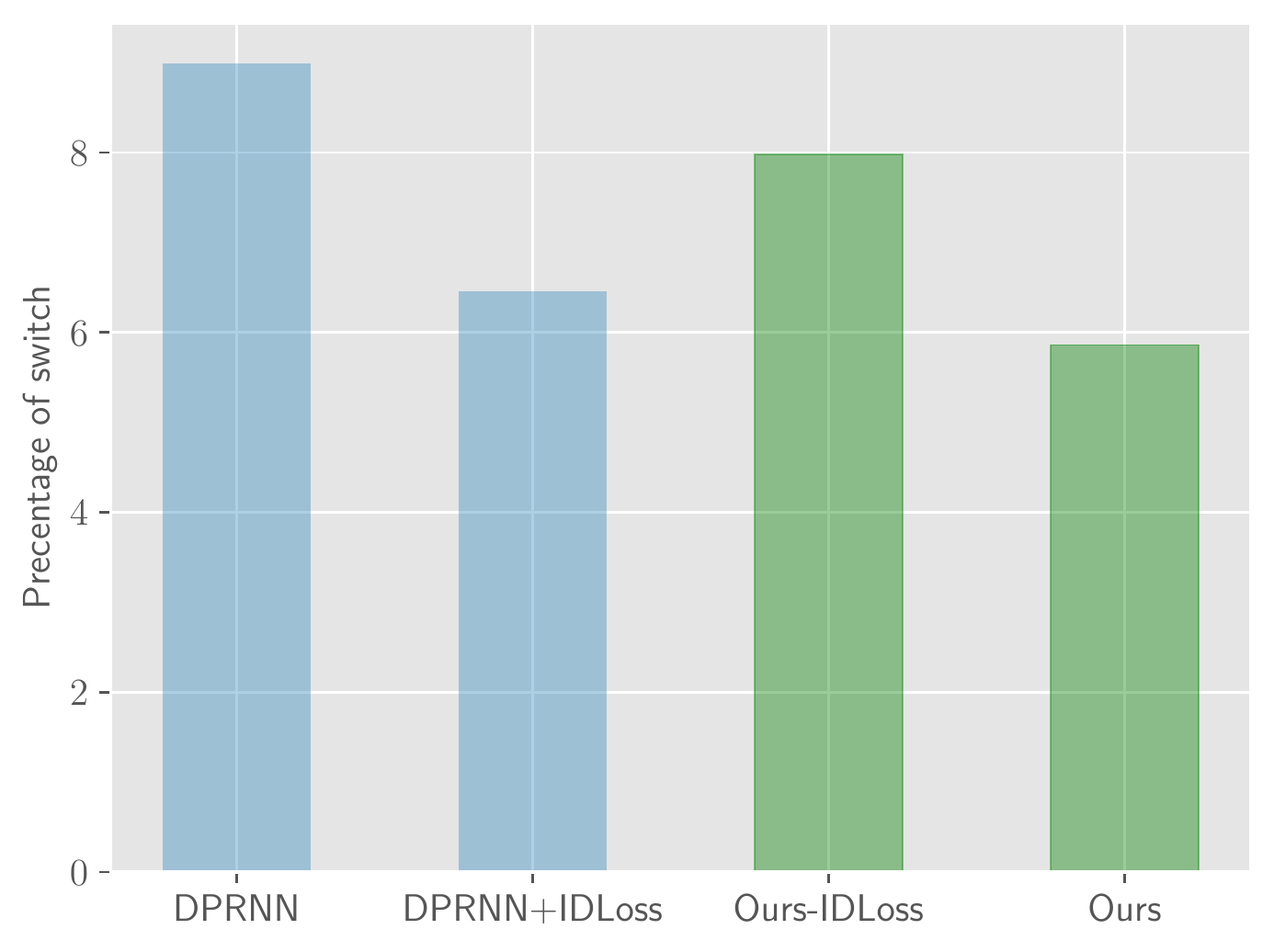}
    \caption{The fraction of samples in which the model produces output channels with an identity switch, using the dataset of two speakers.}
 \label{fig:switch}
\end{figure}

The results suggest that both DPRNN and the proposed model benefit from the incorporation of the identity loss. However, this loss does not eliminate the problem completely. The results are depicted in Figure~\ref{fig:switch}. 

\subsection{Noisy and Reverberated Conditions}
Next, we compared the performance of the proposed approach under noisy and reverberated conditions using WHAM!~\cite{wham} and WHAMR!~\cite{whamr} benchmarks. We compared the proposed method to Conv-TasNet, Chimera++~\cite{wang2018alternative}, Learnable filter bank~\cite{pariente2020filterbank}, and DPRNN. SI-SNRi results are presented in Table~\ref{tab:wham_whamr}. It can be seen that the proposed approach is superior to the baseline methods also under noisy and reverberated conditions.

\begin{table}[t]
    \centering
    \caption{Comparison of the proposed approach against several baselines using WHAM! and WHAMR! datasets.}
    \label{tab:wham_whamr}
    \begin{tabular}{lcc}
    	\toprule
		Model&	WHAM! &	WHAMR! \\
		\midrule
		Chimera++ 				&	9.9 &	-\\
		Learnable filter bank 	&	12.9&	-\\
		Conv-TasNet 			&	12.7&	8.3\\
		DPRNN					&	13.9&	10.3\\
		Ours 					&	\bf 15.2&	\bf 12.2\\
		\bottomrule
	\end{tabular}
\end{table}

\begin{table}[t]
    \centering
    \caption{The results of performing test-time augmentation. The x-axis is the number of shifted versions that were averaged, at inference time, to obtain the final output. The y-axis is the SI-SNRi obtained by this process. DPRNN results are obtained by running the published training code.}
    \label{tab:ttaug}
    \renewcommand{\arraystretch}{1.2}
    \begin{tabular}{@{}l@{~~}c@{~}c@{~}c@{~}c@{~}c@{~}c@{~}c@{}}
    \toprule
    	& \multicolumn{7}{c}{Number of augmentations}\\
	\cmidrule{2-8}
Model	& 0  & 3 & 5 & 7 & 10 & 15 & 20 \\
\midrule
DPRNN(2spk)	& 18.08& 18.11& 18.15& 18.18& 18.19& 18.19& \bf 18.21 \\
Ours(2spk)	& 20.12  & 20.16 & 20.24 & 20.26 & 20.29 & 20.3 & \bf 20.31 \\
\midrule
DPRNN(3spk)	& 14.72  & 15.06 & 15.14 & 15.18 & 15.21 & 15.24 & \bf 15.25 \\
Ours(3spk)	& 16.71  & 16.86 & 16.93 & 16.96 & 16.99 & \bf 17.01 & \bf 17.01 \\
\midrule
DPRNN(4spk)	& 10.37  & 10.49 & 10.53 & 10.54 & 10.56 & 10.57 & \bf 10.58 \\
Ours(4spk)	& 12.88  & 12.91 & 13 & 13.04 & 13.05 & \bf 13.11 & \bf 13.11 \\
\midrule
DPRNN(5spk)	& 8.35  & 8.85 & 8.87 & 8.89 & 8.9 & \bf 8.91 & \bf 8.91 \\
Ours(5spk)	& 10.56  & 10.72 & 10.8 & 10.84 & 10.88 & 10.92 & \bf 10.93 \\
\bottomrule
\end{tabular}
\end{table}

\subsection{Test-time augmentation}
We found out that starting the separation at different points in time yields slightly different results. For this purpose, we cut the mixed audio at a certain time point and then concatenate the first part at the end of the second. Performing this multiple times, at random starting points and then averaging the results tends to improve results.

The averaging process is as follows: first, the original starting point is restored by inverting the shifting process. Then, the channels are then matched (using MSE) to a reference set of channels, finding the optimal permutation. In our experiments, we use the separation results of the original mixed signal as the reference signal. The results from all starting points are then averaged.

Table~\ref{tab:ttaug} depicts the results for both our method and DPRNN. Evidently, as the number of random shifts increases, the performance improves. To clarify: in order to allow a direct comparison with the literature, the results reported elsewhere in this paper are obtained without this augmentation.

\begin{table}[t]
    \centering
    \caption{The results of evaluating models with at least the number of required output channels on the datasets where the mixes contain 2,3,4, and 5 speakers. (a) DPRNN (our training using the authors' published code), (b) our method.}
    \label{tab:nummatrix}
    \begin{tabular}{c}
    \begin{tabular}{@{~}lcccc@{~}}
        \toprule
    	& \multicolumn{4}{c}{Num. speakers in mixed sample}\\
    	\cmidrule{2-5}
        DPRNN model	& 2  &	3 &	4 &	5 \\
        \midrule
        2-speaker model&	18.08	& -	& - & -\\
        3-speaker model&	13.47 & 14.7 & -	& -\\
        4-speaker model&	10.77 &	11.96 &	10.88 &	-\\
        5-speaker model&	7.62 &	9.76 &	9.48 &	8.65\\
        \bottomrule
    \end{tabular} \\
    \vspace{0.1cm}
    (a)\\
    \vspace{0.1cm}
    \begin{tabular}{@{~}lcccc@{~}}
        \toprule
    	& \multicolumn{4}{c}{Num. speakers in mixed sample}\\
    	\cmidrule{2-5}
        Our model	& 2  &	3 &	4 &	5 \\
        \midrule
        2-speaker model&	20.12	&-	&-	&-\\
        3-speaker model&	15.63 & 16.85	&-	&-\\
        4-speaker model&	13.25&	13.46 &	12.88&	-\\
        5-speaker model&	11.02&	11.81 &	11.21 &	10.56\\
        \bottomrule
    \end{tabular}\\
    \vspace{0.1cm}
     (b)\\
     \vspace{-0.5cm}
\end{tabular}
\end{table}

\begin{table*}[t]
    \centering
    \caption{Results of automatically selecting the number of speakers $C$ for a mixed sample $\vx$. Shown are both the confusion matrix and the SI-SNRi results obtained using automatic model selection, in comparison to the results obtained when the number of speakers in the mixture is given. (a) DPRNN, (b) Our model.} 
    \label{tab:multispeaker}
    \begin{tabular}{cc}
	\begin{tabular}{@{~}lcccc@{~}}
    \toprule
	& \multicolumn{4}{c}{Num. speakers in mixed sample}\\
	\cmidrule{2-5}
DPRNN model	& 2  &	3 &	4 &	5 \\
\midrule
2spk& \bf81.3\%	&	   7.9\%			&	3.2\%		&	0.7\%\\
3spk&	15.9\%	&      \bf64.4\%		&	9.9\%		&	2.4\%\\
4spk&	0.7\%	&	   14.5\%		&	\bf 46.2\%	&	11.3\%\\
5spk&	2.1\%	&	   13.2\%		&	40.7\%		&	\bf 85.6\%\\
\midrule
Ours auto-select &	15.88&	12.28&	9.79&	8.53\\
\midrule
Ours known $C$&	18.21	&14.71	&10.37&	8.65\\
\bottomrule
\end{tabular}
&
    \begin{tabular}{@{~}lcccc@{~}}
    \toprule
	& \multicolumn{4}{c}{Num. speakers in mixed sample}\\
	\cmidrule{2-5}
Our model	& 2  &	3 &	4 &	5 \\
\midrule
2spk&	\bf 84.6\%	&	 	3.6\%	  &	1.2\%		&	0.3\%\\
3spk&	13.7\%		&	    \bf 69.0\%&	7.4\%		&	1.6\%\\
4spk&	0.5\%		&	    18.2\%    &	\bf 47.5\%	&	5.8\%\\
5spk&	1.2\%		&	    9.2\%	  &	43.9\%		&	\bf 92.3\%\\
\midrule
Ours auto-select &	18.63&	14.62&	11.48&	10.37\\
\midrule
Ours known $C$&	20.12	&16.85	&12.88&	10.56\\
\bottomrule
\end{tabular}\\
(a)&(b)\\
\end{tabular}
\vspace{0.1cm}
\end{table*}

\subsection{Dealing with an Unknown Number of Speakers}
When there are $C$ speakers in a given mixed audio $x$, one may employ a model that was trained on $C'>C$ speakers. In this case, the superfluous channels seem to produce relatively silent signals for both our method and DPRNN. One can then match the $C'$ output channels to the $C$ channels in the optimal way, discarding $C'-C$ channels, and compute the SI-SNRi score. Tab.~\ref{tab:nummatrix} depicts the results for DPRNN and our method.  As can be seen, the level of results obtained is the same level obtained by the $C'$ model when applied to $C'$ speakers, or slightly better (the mixture audio is less confusing if there are less speakers).

We next apply our model selection method (Section~\ref{sec:select}), which automatically selects the most appropriate model, based on an activity detector algorithm. We consider a silence channel if the output of the activity detection algorithm is above a predefined threshold. For a fair comparison we calibrated the threshold for silence detection to each method separately. We evaluate the proposed method, using a confusion matrix, whether this unlearned method is effective in accurately estimating the number of speakers. Additionally, we measure the obtained SI-SNRi when using the selected model and compare it to the oracle (known number of speaker in the recording).

As can be seen in Table~\ref{tab:multispeaker}, simply by looking for silent output channels, we are able to identify the number of speakers in a large portion of the cases, while maintaining the SI-SNRi values close to the oracle performance. 

\begin{table*}[t]
    \centering
    \caption{Results for the music source separation task. We report SDR results (the higher the better) for the proposed model and several baseline models. All results are reported on the MusDB benchmarks.}
    \label{tab:music}
    \begin{tabular}{lcccccc}
    \toprule
    Model & Wav & \textit{All} & \textit{Drums} & \textit{Bass} & \textit{Other} & \textit{Vocals} \\
    \midrule
 	Open-Unmix & X & 5.33 & 5.73 & 5.23 & 4.02 & 6.32 \\
 	Wave-U-Net & V & 3.23 & 4.22 & 3.21 & 2.25 & 3.25 \\ 
 	Demucs 	   & V & 5.58 & 6.08 & 5.83 & 4.12 & 6.29 \\   
 	Ours 	   & V & \bf{5.82} & \bf{6.15} & \bf{5.88} & \bf{4.32} & \bf{6.92} \\   
	\midrule
	IRM (Oracle) & X & 8.22 & 8.45 & 4.12 & 7.85 & 9.43 \\
\bottomrule
\end{tabular}
\end{table*}

\subsection{Music Source Separation}
Lastly, in order to demonstrate the applicability of the proposed model for other separation tasks, we evaluated our model on the task of music source separation~\footnote{we omitted $\ell_{ID}$ due to its un-relevance to the music source separation task}. In music source separation we are provided with an input mixture, and our goal is to learn a function which outputs C estimated channels, each for a different instrument. Both mixture and the separated channels can be either mono or stereo recordings. In the following experiments we consider the stereo setup since it is more common in modern music recordings.

We evaluate the proposed method on the MusDB dataset~\cite{rafii2017musdb18}. It is comprised of 150 songs sampled at 44100Hz. For each song, we are provided with the separated channels as supervision, where the mixture, is the sum of those four parts. We use the first 84 songs for the training set, the next 16 songs for validation set (we follow the same split as defined in the musdb python package) while the remaining 50 songs are used for test set.

We follow the SiSec Mus evaluation campaign for music separation~\cite{stoter20182018}, where we separate the mixtures into the following four categories: (1) drums, (2) bass, (3) other, and (4) vocals. Unlike the blind-source-separation task, since we know the identity of the target channels (i.e., drums, bass, vocals, other), we do not need to use the permutation invariant training and can directly optimize the L1 distance between the target and output channels. 

We compared the proposed model to several highly competitive baselines; namely Demucs~\cite{defossez2019music}, Wave-U-Net~\cite{stoller2018wave}, and Open-Unmix~\cite{stoter19}. We additionally provide an oracle results of the Ideal Ratio Mask (IRM). We report the median over all tracks of the median Signal-to-Distortion-Ration (SDR) over each track, as done in the SiSec Mus evaluation campaign~\cite{stoter20182018}. For easier comparison, the \textit{All} column is obtained by concatenating the metrics from all sources and then taking the median. 

\noindent{\bf Implementation details.} We tuned all hyper parameters using the validation set. The input kernel size $L$ was set to $14$ and the number of the filter in the preliminary convolutional layer was set to $128$. Similarly to the voice separation task with use $6$ blocks of \ours for the separation module, where each LSTM layer contains $128$ neurons. We optimize the model using Adam optimizer~\cite{kingma2014adam} using a learning rate of $2.5e-4$ and a batch size of $4$.

\noindent{\bf Results.} Results are summarized in Table~\ref{tab:music}. Notice, the proposed model reach the best overall SDR and provide a new state-of-the-art performance~\footnote{We are considering the case of no additional training data. We trained all models using the Demucs package~\citep{defossez2019music}.}. Interestingly, while the proposed method improve performance over all categories, the biggest improvement is on the \textit{Vocals} category. This implies that the proposed model is more suitable for extracting human-speech data.

\section{Related Work}

Single channel speech separation is one of the fundamental problems in speech and audio processing. It was extensively explored over the years~\cite{logeshwari2012survey,martin2018single,ernst2018speech}, where traditionally signal processing techniques were applied to the task~\cite{choi2005blind}. Additionally, most previous research was conducted at the spectrum level of the input signal~\cite{wang2018supervised,ernst2018speech}. 

Recently, deep learning models were suggested for the speech separation task, and have managed to vastly improve the performance over previous methods. \citet{hershey2016deep} suggests a clustering methods in which trained speech embeddings are being used for separation. \citet{yu2017permutation} proposed the Permutation Invariant Training (PIT) over the frame level for source separation, while~\citet{kolbaek2017multitalker} continue this line of work by suggesting the utterance level Permutation Invariant Training (uPIT). \citet{liu2019divide} suggest to use both PIT and clustering over the frame levels for speaker tracking. An excellent survey of deep learning based source separation methods was provided by~\citet{wang2018supervised}. A phase-sensitive objective function with an LSTM neural network was presented in \citet{erdogan2015phase}, showing an improvement of SDR over the CHiME-2 \cite{vincent2013second} dataset.

\citet{wang2019deep} introduced a method to reconstruct the phase in the Fourier transform domain, by determined uniquely the absolute phase between two speakers based on their mixture signal. \citet{wang2018end} propose a novel method to separate speech signals in the time-frequency domain. The method use the magnitude and phase simultaneously in order to separate one speaker from the other. 

An influential method was introduced by~\citet{luo2018tasnet}, where a deep learning method for speech separation over the time domain was presented. It employs thee components: an encoder, a separator and a decoder. Specifically they used a convolutional layer as the encoder, bidirectional LSTMs as the separator network, and a fully connected layer as the decoder that constructs the separated speech signals. Then,~\citet{luo2019conv} suggested to replace the separator network from LSTM to a fully convolutional model using block of time depth separable dilated convolution (Conv-Tasnet). \citet{zeghidour2020wavesplit} propose to infer a set of speaker representations via clustering to better separate the speakers.

Recently, Conv-Tasnet was scaled by~\citet{zhang2020furcanext}, who proposed to train several separator networks in parallel to perform an ensemble. Dual Path RNN blocks were introduced by \citet{luo2019dual}. Such blocks, which we also use, first reorder the encoded representation and then process it across different dimensions. 

Related to our goal of improving the performance on multiple speakers, \citet{takahashi2019recursive} introduce a recursive method for speaker separation, based on the Conv-Tasnet model. The authors suggested to separate out a single speaker at a time in a recursive manner. It is shown that a model that was trained to separate two and three speakers can separate four speakers. 

Another line of work to note is studies which leverages speaker information during separation. In \cite{zmolikova2017speaker,delcroix2018single} a neural network employed estimated i-vectors~\cite{ivectors} in order to estimate masks, which extract the target speaker by generating beamformer filters, while \citet{wang2018voicefilter} proposed to use d-vectors~\cite{dvectors} as speaker embeddings and directly output the target separated sources. 

A clustering method for speech separation was introduced by~\citet{isik2016single, 8045733}. ~\citet{isik2016single} introduced the deep unsupervised separation model. This method estimate the mask by extracting embedding to each segment of the spectrogram and clustering these. A deep attractor network that extract centroids in the high dimensional embedding spaces, in order to obtain the time-frequency bins for each speaker was presented in \cite{chen2017deep}.

 In~\cite{wang2018alternative} multiple clustering network approaches were evaluated and a novel chimera network, which combines mask-inference networks with deep clustering networks, obtains an improvement of 0.7 dB on the WSJ0-2mix dataset over the alternative methods.

Further works on speech separation include input with multiple channels (i.e. multiple microphones). In~\cite{markovich2009multichannel}, the authors present  an extension to the minimum variance distortionless response (MVDR) beamformer~\cite{capon1969high,frost1972algorithm}. The method includes the linearly constrained minimum variance (LCMV) beamformer~\cite{laufer2020global} which is designed to obtain the desired signals and to mitigate the interferences. The MVDR neural beamformer introduce in~\cite{xiao2016study}, predict a mask in time-frequency space, which is then used to estimate the MVDR beam, 
showing an improvement over both the traditional MVDR method and delay-and-sum beamforming.

The FaSNet (filter-and-sum network) method~\cite{luo2019fasnet} includes two-stage processing units. The first learns frame-level time-domain adaptive beamforming filters for a selected reference channel, and second stage calculates the filters for the remaining channels. 
FaSNet improves the MVDR baseline results with $14.3\%$ relative word error rate reduction (RWERR). FaSNet was extended to form the transform-average-concatenate (TAC) model~\cite{luo2019end}. This method employs a transformation module that extract feature to each channel, a copy of these features undergoes global pooling. A concatenation module is then applied to the output of the first and the second modules. This method shows improvement in the separation for noisy environment, and experiments are conducted  with both a varying numbers of microphones and a fix geometry array configuration.

\section{Conclusions}
From a broad perceptual perspective, the cocktail party problem is a difficult instance segmentation problem with many occluding instances. The instances cannot be separated due to continuity alone, since speech signals contain silent parts, calling for the use of an identification-based constancy loss. In this work, we add this component and also use it in order to detect the number of instances in the mixed signal, which is a capability that is missing in the current literature.

Unlike previous work, in which the performance degrades rapidly as the number of speakers increases, even for a known number of speakers, our work provides a practical solution. This is achieved by introducing a new recurrent block, which combines two bi-directional RNNs and a skip connection, the use of multiple losses, and a voice constancy term mentioned above. The obtained results are better than all existing method, in a rapidly evolving research domain, by a sizable gap.

\section*{Acknowledgement} 
The contribution of Eliya Nachmani is part of a Ph.D. thesis research conducted
at Tel Aviv University. We would like to thank anonymous reviewer 4 for suggesting an improvement of the model selection method. The method presented in Sec.~3.3 of this version is the improved one. We also would like to thanks Alexandre D\'efossez for the help with the music source separation experiments.

\clearpage
\bibliography{icml.bib}
\bibliographystyle{icml2020}

\end{document}